\documentclass[12pt]{article}

\usepackage{graphics}
\usepackage{color}
\usepackage{listings}
\usepackage{fullpage}
\usepackage[dvips]{lscape}
\usepackage[dvips]{graphics}
\usepackage{graphicx}
\usepackage{epsf}
\usepackage{epsfig}
\usepackage{rotating}
\usepackage{subfigure}
\usepackage{amsmath}
\usepackage{amssymb}
\usepackage{amsfonts}
\usepackage{mathrsfs}
\usepackage{latexsym}
\usepackage[export]{adjustbox}
\usepackage{authblk}
\usepackage{textpos}

\numberwithin{equation}{section}
\numberwithin{equation}{subsection}
\DeclareMathAlphabet{\mathpzc}{OT1}{pzc}{m}{it}

\begin{document}

\title{Entanglement Entropy of the Gross-Neveu Model \\[30pt]}

\author[1]{Jordan S. Cotler}
\affil[1]{Stanford Institute for Theoretical Physics, Department of Physics, Stanford University}

\author[2]{Mark T. Mueller}
\affil[2]{Center for Theoretical Physics, Department of Physics, Massachusetts Institute of Technology}

\renewcommand\Authands{ and }
\maketitle

\begin{textblock*}{5cm}(11cm,-12.2cm)
\fbox{\footnotesize MIT-CTP-4743}
\end{textblock*}

\begin{abstract}
\noindent
We compute a variational approximation to the entanglement entropy for states of the Gross-Neveu model.  Further, we examine the functional dependence of the entanglement entropy on the coupling and number of colors in the theory.  Our results display non-peturbative behavior.  It is shown that the entanglement entropy is monotonically decreasing and convex with respect to the coupling.   We also show how the behavior of the entanglement entropy under renormalization group transformations is related to the beta function of the Gross-Neveu model, and compare these renormalization group results to our previous work on interacting scalar field theories.

\end{abstract}

\newpage
\tableofcontents

\newpage
\section{Introduction}
\label{sec:introduction}
In previous work, we used the Schr\"{o}dinger picture variational principle to calculate variational approximations to the entanglement entropies of states in interacting scalar field theories \cite{CotlerMueller1}.  These approximations enabled non-perturbative analysis of coupling dependence of the entanglement entropy for $\phi^4$ theory in various dimensions, and have allowed us to study the renormalization group flow of the asymptotically free ``precarious $\phi^4$ theory."  There are no other known analytic techniques which access this non-perturbative information of the entanglement entropy for generic coupled field theories, since for such theories we cannot use the tools of conformal field theory or the Ryu-Takayanagi formalism \cite{Sorkin}--\cite{RyuTakayanagi1}.

In this paper, we turn our attention to fermions, and apply variational techniques to study entanglement entropy in the Gross-Neveu model -- a particularly interesting fermionic field theory with asymptotic freedom \cite{GrossNeveu1}.  The paper is structured as follows:  In Section 2, we introduce the variational method for fermionic field theories in the Schr\"{o}dinger picture, using a formalism developed by Jackiw and Floreanini \cite{JackiwFloreanini,JackiwReview}.  The use of this formalism is illustrated for a free Dirac field, where we also develop techniques for computing the Gaussian effective potential and solving the variational equations.  In Section 3, we apply the machinery of the variational principle with the Grassmann Gaussian ansatz to the Gross-Neveu model.  The variational equations and Gaussian effective potential are computed, and renormalization is discussed.  Finally, in Section 4, we compute the entanglement entropy of the variationally determined Gaussian wavefunctional using standard techniques \cite{Solo1}.  We analyze the dependence of entanglement entropy on the renormalized coupling and number of colors in the theory, and find results that have a distinctly non-perturbative character.   Additionally, we calculate the renormalization group flow for the entanglement entropy, and examine its properties.

\section{Schr\"odinger Picture and Variational Methods for Fermions}
\label{sec:spvm}

\subsection{Formalism}

In this section we introduce the essential aspects of the Schr\"odinger picture and variational methods for interacting quantum field theories with fermions.  These methods were introduced by Jackiw and Floreanini \cite{JackiwFloreanini, JackiwReview}, who developed the formalism and applied it to a variety of problems in quantum field theory.  Several groups have applied variational methods to the study of the Gross-Neveu model, including Latorre and Soto \cite{LatorreSoto} and Kovner and Rosenstein \cite{KovnerRosenstein}, as well as J.H. Yee and collaborators \cite{YeeReview, YeeGN1, YeeGN2}, who utilize the Jackiw-Floreanini formalism.  The main results of this section and the next are intended as a brief review of these variational methods, and the reader is referred to the literature just cited for more details.

Our focus in this paper is on the Gross-Neveu model, which is renormalizable in spatial dimension $d = 1$.  In the large-$N$ limit, where $N$ is the number of different types of fermions (which we refer to as `colors'), it is known \cite{RosensteinWarrPark} that the Gross-Neveu model is also renormalizable in spatial dimension $d = 2$.  In developing the formalism in this section, we will keep the spatial dimension $d$ arbitrary.  The details of the solutions of the variational equations we will need require properties of spinors and Clifford algebras that hold only for both $d = 1$ or $d=2$.  The variational formalism extends in a straightforward way to higher dimensions, but we will not pursue that here.  Details on our conventions for Dirac spinors and Clifford algebras appear in an appendix.  Ultimately, our main interest is in the $d = 1$ version of the Gross-Neveu model, and although we will at times consider the large-$N$ limit, our results are valid for arbitrary $N$, within the scope of validity of the variational approximation.  

The canonical anti-commutation relations for the fermionic Schr\"odinger picture quantum field operators
\begin{align}
	\{ \psi_{\alpha}^a(\mathbf{x}), \psi_{\beta}^b(\mathbf{y}) \} &= 0  \\
	\{ \psi_{\alpha}^a(\mathbf{x}), \psi_{\beta}^{b \dagger}(\mathbf{y}) \} 
	&= \delta^{ab} \, \delta_{\alpha\beta} \, \delta^d(\mathbf{x} - \mathbf{y})
\end{align}
will be realized by Grassmann-valued differential operators acting on a Hilbert space of wavefunctionals of the Grassmann variables $\theta$ and $\theta^{\dagger}$, which carry spinor and color indices and satisfy the anti-commutation relations defining the Grassmann algebra
\begin{align}
	\{ \theta_{\alpha}^a(\mathbf{x}) &, \theta_{\beta}^b(\mathbf{y}) \} = 0  \\
	\{ \theta_{\alpha}^{a \dagger}(\mathbf{x}) &, \theta_{\beta}^{b \dagger}(\mathbf{y}) \} = 0  \\
	\{ \theta_{\alpha}^a(\mathbf{x}) &, \theta_{\beta}^{b \dagger}(\mathbf{y}) \} = 0
\end{align}
Every state $| \Psi \rangle$ of the quantum field theory is represented by a wavefunctional $\Psi[\theta, \theta^{\dagger}]$, and the action of the Schr\"odinger picture field operators on these wavefunctionals is given by the differential operators
\begin{align}
	\psi_{\alpha}^a(\mathbf{x}) 
	&= \frac{1}{\sqrt{2}} \left( \theta_{\alpha}^a(\mathbf{x}) 
	+ \frac{\delta}{\delta \theta_{\alpha}^{a \dagger}(\mathbf{x})}  \right)  \\
	\psi_{\alpha}^{a \dagger}(\mathbf{x}) 
	&= \frac{1}{\sqrt{2}} \left( \theta_{\alpha}^{a \dagger}(\mathbf{x}) 
	+ \frac{\delta}{\delta \theta_{\alpha}^a(\mathbf{x})}  \right)
\end{align}
which satisfy the canonical anti-commutation relations.  In order to represent inner products and operator matrix elements as Grassmann functional integrals, Jackiw and Floreanini describe how to define the dual wavefunctional $\overline{\Psi}(\theta, \theta^{\dagger})$ representing the dual state $\langle \Psi |$.  \footnote{Jackiw and Floreanini use $\Psi^*$ to denote the dual wave functional whereas Yee and collaborators use $\overline{\Psi}$ to denote the same quantity.  We adopt the latter convention.}  This duality is identical to the duality used in the exterior algebra of differential forms, but in an infinite-dimensional setting appropriate for quantum field theory.  The inner product of two states $| \Psi_1 \rangle$ and $| \Psi_2 \rangle$ is then given by the Grassmann functional integral over wavefunctionals
\begin{equation}
	\langle \Psi_2 | \Psi_1 \rangle = \int D\theta \int D\theta^{\dagger} \
	\overline{\Psi}_2 [\theta, \theta^{\dagger}] \Psi_1 [\theta, \theta^{\dagger}]
\end{equation}
Particularly relevant for us will be the Grassmann Gaussian wave functionals, which will serve as the states for our variational ansatz
\begin{equation}
\label{Gaussian1}
	\Psi_G[\theta, \theta^{\dagger}] = 
	N_G \exp \left[ \frac{1}{2} \int D\theta \int D\theta^{\dagger} \ \theta^{\dagger} G \theta \, \right]
\end{equation}
where we are employing a condensed notation; for example
\begin{equation}
	\theta^{\dagger} G \theta = \int d^d x \int d^d y \
	\theta_{\alpha}^{a \dagger}(\mathbf{x}) \, G^{ab}_{\alpha\beta}(\mathbf{x},\mathbf{y}) \, \theta_{\beta}^{b}(\mathbf{y})
\end{equation}
and $G = G^{ab}_{\alpha\beta}(\mathbf{x},\mathbf{y})$ is a totally antisymmetric kernel which may be identified with the equal time Green function.  $N_G$ is a normalization factor which will not play a role in the computation of operator matrix elements.  The dual wavefunctional of the Gaussian in Eqn. (\ref{Gaussian1}) is also a Gaussian
\begin{equation}
	\overline{\Psi}_G[\theta, \theta^{\dagger}] = 
	N_{\bar{G}} \exp \left[ \frac{1}{2} \int D\theta \int D\theta^{\dagger} \ \theta^{\dagger} \overline{G} \theta \right]
\end{equation}
with the dual kernel defined to be $\overline{G} = (G^{\dagger})^{-1}$.  

Within this framework, it is straightforward to compute the Grassmann functional integrals for matrix elements of various fermionic quantum field operators in the Gaussian variational state.  We will not repeat these calculations here; see for example Yee's review article \cite{YeeReview}.  For the matrix elements of operators quadratic in the fields, we have
\begin{equation}
	\langle \Psi_G | \psi_{\alpha}^{a \dagger}(\mathbf{x}) \psi_{\beta}^b(\mathbf{y}) | \Psi_G \rangle
	= W_{\beta\alpha}^{ba}(\mathbf{y}, \mathbf{x})
\end{equation}
or, using a condensed notation combining spatial coordinates, spinor indices, and color indices $A = \{ \mathbf{x}, \alpha, a\}$, we may express the expectation value of the fermion bilinears as
\begin{equation}
	\langle \Psi_G | \psi_A^{\dagger} \psi_B | \Psi_G \rangle = W_{BA}
\end{equation}
The quantity $W$ is defined, in an operator notation with all indices suppressed, by
\begin{equation}  \label{eq:W_definition}
	W = \frac{1}{2} (1 + G)(G + \overline{G})^{-1} (1 +  \overline{G})
\end{equation}
When we look at the Gross-Neveu model in a subsequent section, we will also need matrix elements quartic in the field operators, which will be discussed in that section.  For now, we have what we need to consider the case of a free massive Dirac field.

\subsection{Free Dirac Field}

We will begin our computations in Schr\"odinger picture quantum field theory with Grassmann variables by first considering the free Dirac field.  This allows us to show how the formalism works in a simpler setting in preparation for the discussion in the next section of the Gross-Neveu model with four-fermion interactions.  The Lagrangian density for the free Dirac field with mass $M$ is
\begin{equation}
	\label{eq:Dirac_Lagrangian}
	\mathcal{L} = \overline{\psi} \, (i \gamma^{\mu} \partial_{\mu} - M) \psi
\end{equation}
with corresponding Hamiltonian operator
\begin{equation}
	\label{eq:Dirac_Hamiltonian}
	H = \int d^d x \ \psi^{\dagger}(\mathbf{x}) (- i \boldsymbol{\alpha} \cdot \nabla + \beta M) \psi(\mathbf{x})
\end{equation}
We can express the Dirac Hamiltonian in condensed form as
\begin{equation}
	H = \int d^d x \int d^d y \ \psi^{a \dagger}_{\alpha}(\mathbf{x}) \, 
		h^{ab}_{\alpha\beta}(\mathbf{x},\mathbf{y}) \, \psi_{\beta}^b(\mathbf{y})
	= \psi^{\dagger}_A \, h_{AB} \, \psi_B
\end{equation}
where the Dirac kernel $h$ is defined by
\begin{equation}
	\label{eq:Dirac_Hamiltonian_kernel}
	h_{AB} = h^{ab}_{\alpha\beta}(\mathbf{x},\mathbf{y}) = 
	\delta^{ab} \, (- i \, \boldsymbol{\alpha} \cdot \nabla_{\mathbf{x}} + \beta M)_{\alpha\beta} \, \delta^d(\mathbf{x} - \mathbf{y})
\end{equation}
The expectation value of the Dirac Hamiltonian in the Gaussian variational state $| \Psi_G \rangle$ involves only the two-fermion matrix elements, given in the preceding section.
\begin{equation}
	\langle \Psi_G | H | \Psi_G \rangle
	= \langle \Psi_G | \, \psi_A^{\dagger} \, h_{AB} \, \psi_B | \Psi_G \rangle 
	= h_{AB} W_{BA}
	= \text{Tr} (h W)
\end{equation}
The trace denoted by ``Tr'' is over all indices in the condensed notation, including spatial coordinates, and $W$ was defined in \eqref{eq:W_definition}.  It is useful to formulate our kernels in momentum space, so we take the Fourier transforms
\begin{align}
	h_{\alpha\beta}^{ab}(\mathbf{x},\mathbf{y}) 
	&= \int \frac{d^d p}{(2\pi)^d} \, e^{i \mathbf{p} \cdot (\mathbf{x} - \mathbf{y})} h_{\alpha\beta}^{ab}(\mathbf{p})  \\[4pt]
	W_{\alpha\beta}^{ab}(\mathbf{x},\mathbf{y}) 
	&= \int \frac{d^d p}{(2\pi)^d} \, e^{i \mathbf{p} \cdot (\mathbf{x} - \mathbf{y})} W_{\alpha\beta}^{ab}(\mathbf{p})
\end{align}
with the Dirac Hamiltonian kernel in momentum space given by
\begin{equation}
	h(\mathbf{p}) = \boldsymbol{\alpha} \cdot \mathbf{p} + \beta M
\end{equation}
The result may be written as
\begin{equation}
	\langle \Psi_G | H | \Psi_G \rangle = \text{Tr} (h W) = \int d^d x \, V_{GE}
\end{equation}
where $V_{GE}$, the Gaussian effective potential, is given by
\begin{equation}
	V_{GE} = \int \frac{d^d p}{(2\pi)^d}  \ \text{tr} \, [ h(\mathbf{p}) W(\mathbf{p}) ]
\end{equation}
The trace denoted by the lower-case ``tr'' is now over only the spinor and color indices, the trace over spatial coordinates having been handled in passing to momentum space.  Using the momentum space Green functions defined by
\begin{align}
	G_{\alpha\beta}^{\, ab}(\mathbf{x},\mathbf{y}) 
	&= \int \frac{d^d p}{(2\pi)^d} \, e^{i \mathbf{p} \cdot (\mathbf{x} - \mathbf{y})} \, G_{\alpha\beta}^{\, ab}(\mathbf{p})  \\[4pt]
	\overline{G}_{\alpha\beta}^{\, ab}(\mathbf{x},\mathbf{y}) 
	&= \int \frac{d^d p}{(2\pi)^d} \, e^{i \mathbf{p} \cdot (\mathbf{x} - \mathbf{y})} \, \overline{G}_{\alpha\beta}^{\, ab}(\mathbf{p})
\end{align}
the quantity $W$, defined in \eqref{eq:W_definition}, takes the form in momentum space
\begin{equation}
	W(\mathbf{p}) = \frac{1}{2} (1 + G(\mathbf{p}))(G(\mathbf{p}) + \overline{G}(\mathbf{p}))^{-1} (1 +  \overline{G}(\mathbf{p}))
\end{equation}
with spinor and color indices suppressed.  

We now want to vary the effective potential $V_{GE}$, with variations $\delta G$ and $\delta \overline{G}$, which may be treated as independent variations.  A short calculation gives the variation of $W$ in operator form
\begin{equation}
	\label{eq:W_variation}
	\delta W = \frac{1}{2} (1 - \overline{G}) \, \Delta G \, (1 + \overline{G}) + \frac{1}{2} (1 + G) \, \Delta \, \overline{G} \, (1 - G)
\end{equation}
where we have defined
\begin{align}
	\Delta G &= - (G + \overline{G})^{-1} \delta G \, (G + \overline{G})^{-1}  \\
	\Delta \overline{G} &= - (G + \overline{G})^{-1} \delta \overline{G} \, (G + \overline{G})^{-1}
\end{align}
Using this result to compute the variation of the effective action, and using the cyclic property of the trace, we arrive at the variational equations for $G$ and $\overline{G}$
\begin{align}
	\label{eq:G_variational_equation}
	(1 - G) \, h \, (1 + G) &= 0  \\
	\label{eq:Gbar_variational_equation}
	(1 + \overline{G}) \, h \, (1 - \overline{G}) &= 0
\end{align}
We would like to emphasize now that although we have been using the kernel $h$ for the free Dirac Hamiltonian, the variation of $\text{Tr} (h W)$ with respect to $G$ and $\overline{G}$ leads to the above variational equations \eqref{eq:G_variational_equation} and \eqref{eq:Gbar_variational_equation} for \emph{any} hermitian kernel $h$.  This fact will be very useful in the next section when we derive the variational equations for the Gross-Neveu model which is an interacting quantum field theory.  In the remainder of this section we will consider an arbitrary hermitian kernel $h$ in deriving the solution of the variational equations.

Yee and collaborators \cite{YeeReview,YeeGN1,YeeGN2} have developed a method for obtaining all of the solutions to the above matrix equations, which we now describe.  We will consider the first of the pair of variational equations for $G$, Eqn. \eqref{eq:G_variational_equation}, and having found its solution we will see that the other variational equation is also satisfied by the same solution.  Multiplying the first variational equation by $h$ and defining the kernel $K = hG$, the variational equation becomes
\begin{equation}
	h^2 - K^2 + [h,K] = 0
\end{equation}
It is evident that there are always trivial solutions to these equations with $K = h$.  Expanding in a basis of 2$\times$2 matrices given by the identity $1$ and the standard Pauli matrices $\sigma_i$, with $i = 1,2,3$ and an implicit summation convention on $i$,
\begin{align}
	h &= h_0 \, 1 + h_i \, \sigma_i  \\
	K &= K_0 \, 1 + K_i \, \sigma_i
\end{align}
Writing out the variational equation for $K$ in components, the solutions of the variational equation in this basis must satisfy
\begin{align}
	K_i &= \frac{h_0}{K_0} h_i  \\[4pt]
	(K_0^2)^2 \, - \, &(h_0^2 + \omega^2) K_0^2 + h_0^2 \, \omega^2 = 0
\end{align}
where we have defined
\begin{equation}
	\label{eq:omega_definition}
	\omega^2 = h_i h_i
\end{equation}
The above quadratic equation for $K_0^2$ has two solutions: either $K_0^2 = h_0^2$ or $K_0^2 = \omega^2$.  The first choice reproduces the trivial solution mentioned above; the second choice gives the non-trivial solution
\begin{equation}
	K = \pm \frac{1}{\omega} ( \omega^2 \, 1 + h_0 h_i \sigma_i)
\end{equation}
Now, using the matrix inverse of the kernel $h$
\begin{equation}
	h^{-1} = \frac{1}{h_0^2 - \omega^2} (h_0 1 - h_i \sigma_i )
\end{equation}
and $G = h^{-1} K$, we arrive at the solution in momentum space of the variational equation for the Green function $G(\mathbf{p})$ in terms of the Hamiltonian kernel $h(\mathbf{p})$
\begin{equation}
	\label{eq:G_solution}
	G(\mathbf{p}) = - \frac{1}{\omega(\mathbf{p})} \, h_i(\mathbf{p}) \, \sigma_i
\end{equation}
Note that $G_0 = 0$ has no identity matrix component and $G$ has no dependence on the identity matrix component $h_0$ of the hamiltonian kernel $h$.  (For the Dirac Hamiltonian, $h_0 = 0$ has no identity matrix component, but this need not be the case for a general Hamiltonian kernel.)  The choice of the plus or minus sign is associated with the definition of the vacuum state.  The standard choice of filling the negative energy states (Dirac sea) leads to choosing the minus sign in the solution for the Green function $G$.  We will see this explicitly when we compute the effective potential in terms of this solution for $G$.  

Having found this explicit solution for the Green function, we note some important properties.  Because the Pauli matrices are hermitian and the Pauli-components of $h(\mathbf{p})$ are real, we see immediately that $G$ is also hermitian:  $G^{\dagger} = G$.  The inverse of $G$ is also easily computed
\begin{equation}
	G^{-1} = - \omega \frac{1}{-\omega^2} (-h_i \sigma_i) = - \frac{1}{\omega} \, h_i \sigma_i = G
\end{equation}
so we find that
\begin{equation}
	\overline{G} = (G^{\dagger})^{-1} = G
\end{equation}
This confirms that both of the variational equations which resulted from varying the effective potential with respect to $G$ and $\overline{G}$ are satisfied.  Furthermore, using $\overline{G} = G^{-1} = G$, the quantity $W$ appearing in expectation values of the fermion field operators in the Gaussian state simplifies to
\begin{equation}
	\label{eq:W_simplified}
	W = \frac{1}{2} (1 + G)(G + \overline{G})^{-1} (1 + \overline{G}) = \frac{1}{2} (1 + G)
\end{equation}

Returning now to the specific Dirac form of the Hamiltonian kernel $h(\mathbf{p})$, and using the Dirac representation we have chosen (see appendix) relating the $\alpha$ and $\beta$ matrices to the Pauli matrices, we have the Pauli components of $h(\mathbf{p})$
\begin{equation}
	h_0(\mathbf{p}) = 0  \quad\quad
	h_m(\mathbf{p}) = \mathbf{p}_m  \quad\quad
	h_3(\mathbf{p}) = M
\end{equation}
where $m$, the spatial index, takes only the value 1 in 1+1 dimensions and the values 1,2 in 2+1 dimensions.  Using the definition of $\omega$ given in Eqn. \eqref{eq:omega_definition}, we find that
\begin{equation}
	\omega^2(\mathbf{p}) = \mathbf{p}^2 + M^2
\end{equation}
Therefore, the Green function in momentum space for the free Dirac theory is
\begin{equation}
	G(\mathbf{p}) = \frac{-1}{\sqrt{\mathbf{p}^2 + M^2}} \, (\boldsymbol{\alpha} \cdot \mathbf{p} + \beta M)
\end{equation}
The Gaussian effective potential in the free Dirac theory takes the physically intuitive and simple form
\begin{equation}
	\label{eq:free_V_GE}
	V_{GE} = \int \frac{d^d p}{(2\pi)^d}  \ \text{tr} \, [ h(\mathbf{p}) W(\mathbf{p}) ]
	= - N \int \frac{d^d p}{(2\pi)^d} \, \sqrt{\mathbf{p}^2 + M^2}
\end{equation}
where we have used the explicit form of the Dirac Hamiltonian, the simplified form of the solution for $W$ given in Eqn. \eqref{eq:W_simplified}, and the fact that the $\alpha$ and $\beta$ matrices (or their Pauli matrix representations) have vanishing traces.  The overall factor of $N$ arises simply because of the trace over the color indices.  As mentioned above, the choice of the minus sign in the solution for the Green function $G$ has led us directly to the minus sign for the Gaussian effective potential for the free Dirac Hamiltonian, with its interpretation as a filling of the Dirac sea to define the vacuum state of the free Dirac theory.

\section{Variational Calculations for the Gross-Neveu Model}
\label{sec:gnm}

The Gross-Neveu model \cite{GrossNeveu1} is described by the Lagrangian density
\begin{equation}
	\mathcal{L} = \overline{\psi} \, i \gamma^{\mu} \partial_{\mu} \psi + \frac{g^2}{2} (\overline{\psi} \psi)^2
\end{equation}
with corresponding Hamiltonian operator
\begin{equation}
	H = H_D + H_I 
	= \int d^d x \, [ - i \psi^{\dagger} \boldsymbol{\alpha} \cdot \nabla \psi - \frac{g^2}{2} (\psi^{\dagger} \beta \psi)^2 ]
\end{equation}
where spinor and color indices have been suppressed.  The first term $H_D$ is the free Dirac Hamiltonian discussed in the previous section, but now with no mass term.  The interaction term $H_I$ is parameterized by the real bare coupling $g$.  

It is well-known \cite{GrossNeveu1}, and we will see using variational methods, that in the Gross-Neveu model a fermion mass is generated dynamically and non-perturbatively even though the Hamiltonian has no bare mass term.  It is also well-known, and again we will see using variational methods, that the Gross-Neveu model is asymptotically free.  These two properties explain much of the interest in this model.  In addition to these intriguing properties, the Gross-Neveu model will serve us well for developing variational calculations of entanglement entropy in interacting quantum field theories with fermions.  Such 1+1-dimensional models may be of direct physical interest, but also serve as a departure point for investigating models such as fermions interacting with gauge fields in 3+1 dimensions.

\subsection{Gap Equation and Effective Potential}
\label{ssec:geep}

Our first task in employing the Gaussian variational method is to compute the expectation value of the Hamiltonian operator in the Grassmann Gaussian variational state.  We have already done this for the free Dirac Hamiltonian, which we will now write, with spinor and color indices suppressed, as
\begin{equation}
	\langle \Psi_G | H_D | \Psi_G \rangle = \int d^d x \ \text{tr} [ (h_D W)(\mathbf{x},\mathbf{x})] 
\end{equation}
where with the usual composition of operator kernels, 
\begin{equation}
	(h_D W)(\mathbf{x},\mathbf{x}) = \int d^d y \ h_D(\mathbf{x},\mathbf{y}) W(\mathbf{y},\mathbf{x})
\end{equation}
The Dirac Hamiltonian kernel $h_D$ is given in Eqn. \eqref{eq:Dirac_Hamiltonian_kernel} with $M = 0$.

To compute the expectation value of the interaction term of the Hamiltonian, we need the general four-fermion matrix element in the Grassmann Gaussian variational state.  Using the condensed index notation, a short calculation \cite{YeeReview} gives
\begin{equation}
	\langle \Psi_G | \psi_A^{\dagger} \, \psi_B \, \psi_C^{\dagger} \, \psi_D | \Psi_G \rangle
	= W_{BA} W_{DC} - W_{DA} W_{BC} + W_{DA} \delta_{BC} 
\end{equation}
Using this result, we have
\begin{align}
	\langle \Psi_G | H_I | &\Psi_G \rangle 
	= - \frac{g^2}{2} \int d^d x \, \beta_{\alpha\beta} \, \delta_{ab} \, \beta_{\gamma\delta} \, \delta_{cd} \,
	\langle \Psi_G | \, \psi^{\dagger}_{\alpha a} \, \psi_{\beta b} \, \psi^{\dagger}_{\gamma c} \, \psi_{\delta d} | \Psi_G \, \rangle  \\
	&= - \frac{g^2}{2} \int d^d x \, [ \, \text{tr} (\beta \tilde{W}) \, \text{tr} (\beta \tilde{W}) 
	- \text{tr}(\beta \, \tilde{W} \beta \, \tilde{W}) + \text{tr} (\beta \, \tilde{\delta} \, \beta \, \tilde{W}) \, ]
\end{align}
where the traces are over both spinor and color indices.  We have defined the position-space-diagonal elements of the kernel $W$ and the $\delta$-function kernel as
\begin{align}
	\tilde{W}_{\alpha\beta}^{ab} &= W_{\alpha\beta}^{ab}(\mathbf{x},\mathbf{x})  \\[4pt]
	\tilde{\delta}_{\alpha\beta}^{ab} &= \delta^{ab} \, \delta_{\alpha\beta} \, \delta^d(\mathbf{x}, \mathbf{x})
\end{align}
These quantities arise because all of the Schr\"odinger picture quantum field operators in the interaction Hamiltonian are evaluated at the same point in space, and are divergent as in any local quantum field theory.  These divergences will be given meaning in the process of renormalization discussed in the next section.

Although we have now computed the expectation value of the full Hamiltonian in the Gaussian variational state $| \Psi_G \rangle$, we are not done.  Unlike the situation with scalar fields, we cannot introduce a variational parameter into the Gaussian variational wavefunctional which is the expectation of a single fermion field and still be consistent with Lorentz invariance.  Instead, following the original treatment of Gross and Neveu \cite{GrossNeveu1} and within the variational framework discussed by Yee and collaborators \cite{YeeGN1,YeeGN2}, we introduce a constraint that allows us to consider the expectation value of a fermion bilinear term.  We define the auxiliary field $\sigma$ by
\begin{equation}
	\sigma = - g \, \langle \Psi_G | \, \overline{\psi} \, \psi \, | \Psi_G \rangle 
	= - g \, \langle \Psi_G | \, \psi^{\dagger} \beta \, \psi \, | \Psi_G \rangle
\end{equation}
together with an associated Lagrange multiplier field $\chi$, and add to the Hamiltonian a constraint term $H_C$ with expectation value
\begin{align}
	\langle \Psi_G | H_C | \Psi_G \rangle
	&= \int d^d x \ \chi \, [ \, \sigma + g \, \langle \Psi_G | \psi^{\dagger} \beta \, \psi | \Psi_G \rangle \, ]  \\
	&= \int d^d x \ \chi \, [ \, \sigma + g \, \text{tr}(\beta \tilde{W}) \, ]
\end{align}
where once again we have used the result for the expectation of bilinears of fermion field operators given in a previous section.  With the constraint included, the expectation value of the full Hamiltonian $H = H_D + H_I + H_C$ may be written in the form
\begin{equation}
\begin{split}
	&\langle \Psi_G | H | \Psi_G \rangle = \langle \Psi_G | H_D + H_I + H_C | \Psi_G \rangle  \\[6pt]
	= \int d^d x [ \, \chi \sigma - \frac{\sigma^2}{2} &+ \text{tr}(h_D W) + g \chi \, \text{tr} (\beta \, \tilde{W})
	+ \frac{g^2}{2} ( \text{tr}(\beta \, \tilde{W} \beta \, \tilde{W}) - \text{tr} (\beta \, \tilde{\delta} \, \beta \, \tilde{W}) ) \, ]
\end{split}
\end{equation}

We are now ready to vary this effective Hamiltonian to get the variational equations.  Varying with respect to $G$ and $\overline{G}$, the $\sigma$ and $\chi$ terms do not contribute, and the variation takes the form
\begin{equation}
	\delta \langle \Psi_G | H | \Psi_G \rangle = \int d^d x \, \text{tr} (h \, \delta W)
\end{equation}
with $\delta W$ given in terms of $\delta G$ and $\delta \overline{G}$ in Eqn. \eqref{eq:W_variation}, and we have now defined the kernel $h$ by
\begin{equation}
	\label{eq:variational_kernel}
	h = h_D + g \chi \beta + \frac{g^2}{2} \beta (2 \tilde{W} - \tilde{\delta}) \beta
\end{equation}
This is the kernel that results from the variation, although it is not the kernel associated with the full Hamiltonian.  It will lead us immediately to the variational equations and their solutions, as discussed in detail in the section on the free Dirac field where we emphasized that those techniques may be applied to more general kernels $h$.  First we note that the full effective Hamiltonian can be expressed in terms of this kernel $h$ as
\begin{equation}
	\label{eq:full_effective_hamiltonian}
	\langle \Psi_G | H | \Psi_G \rangle
	= \int d^d x [ \, \chi \sigma - \frac{\sigma^2}{2} + \text{tr}(h W) - \frac{g^2}{2} \text{tr}(\beta \, \tilde{W} \beta \, \tilde{W} ) \,]
\end{equation}
which we will return to shortly when we compute the Gaussian effective potential.  Recalling Eqn. \eqref{eq:G_variational_equation}, our variational equation for $G$, with $h$ as just defined, is
\begin{equation}
	(1 - G) \, h \, (1 + G) = 0
\end{equation}
We found that the general solution in momentum space for $G(\mathbf{p})$ in the Pauli basis was given in Eqn. \eqref{eq:G_solution} by
\begin{equation}
	G(\mathbf{p}) = \frac{-1}{\sqrt{\omega^2(\mathbf{p})}} \, h_i(\mathbf{p}) \, \sigma_i
\end{equation}
with $\omega^2 = h_i h_i$ originally defined in Eqn. \eqref{eq:omega_definition}.  Using the result that the quantity $W$ simplifies in terms of this solution for $G$ according to Eqn. \eqref{eq:W_simplified}, we have $2 \tilde{W} - \tilde{\delta} = \tilde{G}$, so the kernel $h$ given above in \eqref{eq:variational_kernel} now takes the simple form in momentum space
\begin{equation}	
	h(\mathbf{p}) = \boldsymbol{\alpha} \cdot \mathbf{p} + g \chi \beta + \frac{g^2}{2} \beta \tilde{G} \beta
\end{equation}
We see that $G$ is expressed in terms of $h$, and that $h$ in turn depends on $G$, or more accurately on $\tilde{G}$.  The quantity $\tilde{G}$ is the divergent position-space-diagonal piece of $G$, which may be expressed as an integral over the momentum space Green function $G(\mathbf{p})$ :
\begin{equation}
	\tilde{G} = G(\mathbf{x},\mathbf{x}) = \int \frac{d^d p}{(2\pi)^d} \, G(\mathbf{p})
\end{equation}
To solve our coupled equations for $h$ and $G$ we continue to use the Pauli basis and the Dirac representation for the $\alpha$ and $\beta$ matrices featured in the section on the free Dirac theory, and which are described in more detail in the appendix.  In agreement with Yee and collaborators \cite{YeeGN1,YeeGN2}, we find that necessarily $\tilde{G}_i = 0$ for both the $i = 1$ and $i = 2$ matrix components of $\tilde{G}$ in the Pauli basis.  Only $\tilde{G}_3$ is non-zero, and we are led to the final form for $h$ and $G$
\begin{equation}
	\label{eq:hm}
	h(\mathbf{p}) = \boldsymbol{\alpha} \cdot \mathbf{p} + \beta m
\end{equation}
and
\begin{equation}
	\label{eq:gm}
	G(\mathbf{p}) = \frac{-1}{\sqrt{\mathbf{p}^2 + m^2}} \, (\boldsymbol{\alpha} \cdot \mathbf{p} + \beta m)
\end{equation}
which look nearly identical to the results for the free Dirac theory.  However, instead of the mass parameter $M$ that we included ab initio in the free theory, a new mass parameter $m$ has appeared even though the original Gross-Neveu Hamiltonian did not contain any bare mass term.  This mass parameter $m$ is defined by
\begin{equation}
	m = g \chi + \frac{g^2}{2} \tilde{G}_3
\end{equation}
where 
\begin{equation}
	\tilde{G}_3 = \int \frac{d^d p}{(2\pi)^d} \, G_3(\mathbf{p}) = \int \frac{d^d p}{(2\pi)^d} \, \frac{-m}{\sqrt{\mathbf{p}^2 + m^2}}
\end{equation}
We have arrived at an equation for the mass parameter $m$:
\begin{equation}
	\label{eq:gap_equation}
	m = g \chi - \frac{g^2}{2} \int \frac{d^d p}{(2\pi)^d} \, \frac{m}{\sqrt{\mathbf{p}^2 + m^2}}
\end{equation}
This important equation is known as the \emph{gap equation}.

Before moving on to the discussion of renormalization, we return to the computation of the Gaussian effective potential by inserting the solution for $G$ into the effective Hamiltonian given in Eqn. \eqref{eq:full_effective_hamiltonian}.  In addition to the auxiliary field terms, there are two other terms.  For the first term, arising from $\text{tr}(h W)$, we use the final expressions for $h(\mathbf{p})$ and $G(\mathbf{p})$ given above in Eqn.'s \eqref{eq:hm} and \eqref{eq:gm}, and the trace properties of the Pauli matrices.  A short computation shows that this gives a contribution to the effective potential
\begin{equation}
	\int \frac{d^d p}{(2\pi)^d}  \ \text{tr} \, [ h(\mathbf{p}) W(\mathbf{p}) ]
	= - N \int \frac{d^d p}{(2\pi)^d} \, \sqrt{\mathbf{p}^2 + m^2}
\end{equation}
This result closely resembles our earlier result for the Gaussian effective potential of the free Dirac theory given in Eqn. \eqref{eq:free_V_GE}, but now the mass parameter $m$ which satisfies the gap equation has replaced $M$.  For the second term, a short computation using the Pauli basis and the solution for $W$ in terms of $G$ given by Eqn. \eqref{eq:W_simplified} gives
\begin{equation}
	- \frac{g^2}{2}  \text{tr} ( \beta \, \tilde{W} \beta \, \tilde{W} ) = - \frac{g^2 N}{4} (1 + \tilde{G}_3^2)
	= - \frac{N}{g^2} (m - g \chi)^2 + \cdots
\end{equation}
where we have used the definition of $m$, and where the factor of $N$ results from the trace over the color indices (surviving traces over the spinor indices contribute a multiplicative factor of 2).  The term indicated by $+ \cdots$ is an additive divergent constant, and has been dropped.  In the next section on renormalization, we will discuss our choice for the zero value of the effective potential on physical grounds.  Putting all of these pieces together, we arrive at the result for the Gaussian effective potential, expressed in terms of $\chi$, $\sigma$, and $m$ :
\begin{equation}
	 V_{GE} = \chi \sigma - \frac{\sigma^2}{2} 
	 - N \int \frac{d^d p}{(2\pi)^d} \, \sqrt{\mathbf{p}^2 + m^2} 
	 - \frac{N}{g^2} (m - g \chi)^2
\end{equation}
We may simplify this further by varying with respect to the auxiliary field $\sigma$, which gives $\sigma = \chi$, thus identifying the auxiliary field with the Lagrange multiplier field.  The resulting Gaussian effective potential is
\begin{equation} 
	\label{eqn:vge}
	V_{GE} = \frac{\chi^2}{2} - N \int \frac{d^d p}{(2\pi)^d} \, \sqrt{\mathbf{p}^2 + m^2} - \frac{N}{g^2} (m - g \chi)^2
\end{equation}
Using the gap equation which relates the field $\chi$ to the mass parameter $m$, we may treat the Gaussian effective potential $V_{GE}$ as a function of just $m$, or just $\chi$.

\subsection{Renormalization of the Effective Potential}
\label{ssec:rep}

The structure of the divergences and the results for the renormalized gap equation and Gaussian effective potential take different forms in spatial dimensions $d = 1$ and $d = 2$.  As mentioned earlier, the Gross-Neveu model is renormalizable at large-$N$ in $d = 2$.  Our methods apply to the $d=2$ case, but we choose to focus our discussion on the $d =1$ case.  Our goal is to define a renormalized coupling $g_R$ in terms of the Gaussian effective potential, giving a renormalization prescription relating this renormalized coupling to the bare coupling $g$.  There will also be a renormalization of the field $\chi$.  With these renormalization prescriptions, we will arrive at our final results for the renormalized Gaussian effective potential and renormalized gap equation.  These results will allow us to compute a variational approximation to the entanglement entropy in the next section.

We begin with the definition of the renormalized coupling $g_R$.  Consider the Gaussian effective potential as a function of $m$ alone, as discussed at the end of the previous section, and define $g_R$ by
\begin{equation}
	\frac{1}{g_R^2} = V''_{GE}(m) \vert_{m = \mu}
\end{equation}
where derivatives with respect to $m$ are denoted by a prime, and where $\mu$ is an arbitrary mass scale which we will refer to as the \emph{renormalization scale}.  To compute the necessary derivatives of $V_{GE}$, we will need to compute derivatives of $\chi$ with respect to $m$ using the gap equation \eqref{eq:gap_equation}.  We will write the gap equation in the form
\begin{equation}
	g \chi = m + g^2 m \, I_0(m)
\end{equation}
where $I_0(m)$ is one of a family of integrals defined by
\begin{equation}
	I_n(m) = \int \frac{d^d p}{(2\pi)^d \, 2 \omega(\mathbf{p})} \, \omega(\mathbf{p})^{2n}
\end{equation}
with $\omega(\mathbf{p}) = \sqrt{\mathbf{p}^2 + m^2}$ as usual.  These integrals satisfy various recursion relations and identities for their derivatives with respect to $m$ that streamline the discussion of renormalization, and allow us to proceed without having to utilize any particular form of regularization.  These integrals and their identities were introduced by Stevenson \cite{Stevenson1, Stevenson2} and are discussed there and in our paper on $\phi^4$ theory \cite{CotlerMueller1}.  Computing the first derivative of $\chi$ with respect to $m$, we have
\begin{equation}
	g \chi'(m) = 1 + g^2 I_0(m) + g^2 m I_0'(m)
\end{equation}
In $d = 1$, $I_0(m)$ is logarithmically divergent and $I_0'(m)$ is finite. We will systematically drop terms that go to zero as $g$ goes to zero, which includes the third term above.  Keeping these sub-leading terms in $g$ does not change the final form of the renormalization prescription for the coupling.  With the derivative of $\chi$ with respect to $m$ given above, and using the Gaussian effective potential in the form \eqref{eqn:vge}, computing derivatives of $V_{GE}$ is straightforward.  Evaluating the result for the second derivative at the renormalization mass scale $\mu$ gives
\begin{equation}
	V''_{GE}(\mu) = \frac{1}{g_R^2} = \frac{1}{g^2} [1 - 2(N - 1) g^2 I_0(\mu) + (2N - 1) (g^2 I_0(\mu))^2 ]
\end{equation}
The only way for the renormalized coupling $g_R$ to be finite is for the quantity in square brackets above to go to zero as the momentum space cutoff goes to infinity.  This leads to a quadratic equation for $g^2 I_0(\mu)$, giving two solutions that might lead to a finite renormalized coupling.  Either $g^2 I_0(\mu) = -1$ or
\begin{equation}
	g^2 I_0(\mu) = \frac{1}{2N - 1}
\end{equation}
We choose the latter.  The situation here is discussed in more detail in \cite{LatorreSoto,YeeGN1}, where they explain why the former choice for $g^2$ leads to an apparently inconsistent theory.  In some respects, renormalization within the variational approximation in the Gross-Neveu model in 1+1 dimensions appears to be similar to renormalization in $\phi^4$ theories in 3+1 dimensions, where the two renormalization prescriptions lead to the so-called ``precarious'' and ``autonomous'' versions of the theory \cite{StevensonTarrach}, with evidence that the latter is also not a self-consistent theory.  So, we are led to the coupling renormalization prescription
\begin{equation}
	\label{eq:coupling_renormalization}
	\frac{1}{g_R^2} = \frac{1}{g^2} - (2N - 1) I_0(\mu)
\end{equation}

We can now discuss the renormalized gap equation.  Using the above coupling renormalization, we find
\begin{align}
	g \chi &= m + g^2 m \, I_0(m)  \\[4pt]
	&= m \, \left[1 + g^2 I_0(\mu) + g^2 ( I_0(m) - I_0(\mu) ) \right]  \\[4pt]
	&= m \left[ \, \frac{2N}{2N - 1} - \frac{g^2}{4\pi} \log \frac{m^2}{\mu^2} \right]
\end{align}
but as $g \rightarrow 0$, the second term vanishes and we find the simple result for the renormalized gap equation
\begin{equation}
	\label{eq:renormalized_gap_equation}
	g \chi = g_R \chi_R = \frac{2N}{2N - 1} \, m
\end{equation}
where this equation defines the renormalized field $\chi_R$.

Having derived the renormalization prescription for the coupling and the renormalized form of the gap equation, we turn to the renormalized form of the Gaussian effective potential.  A straightforward calculation shows that with the above coupling and field renormalizations, all of the divergent terms cancel (with the exception of additive constants that are dealt with through our physically-motivated choice of the zero of the Gaussian effective potential), leading to the result
\begin{equation}
	\label{eq:renormalized_gep_m_form}
	V_{GE}(m) = \frac{1}{2} \chi_R^2 - \frac{N}{g_R^2} (m - g_R \chi_R)^2 
	+ \frac{N}{4\pi} \left( m^2 \log \frac{m^2}{\mu^2} - m^2 \right)
\end{equation}
As with the un-renormalized Gaussian effective potential, we can treat $V_{GE}(m)$ as a function of \textit{only} $m$, with $\chi_R$ expressed in terms of $m$ via the renormalized gap equation.  Alternatively, we define a reference value $\chi_0$ in terms of the mass scale $\mu$ by
\begin{equation}
	\label{eq:chi0_definition}
	\chi_0 = \left( \frac{2N}{2N - 1} \right) \frac{\mu}{e g_R}
\end{equation}
in terms of which the renormalized Gaussian effective potential takes the form
\begin{equation}
	\label{eq:renormalized_gep_chi_form}
	V_{GE}(\chi_R) = \frac{1}{2} \left( \frac{2N - 1}{2N} \right) \chi_R^2 
	+ \frac{N}{4\pi} \left( \frac{2N - 1}{2N} \right)^2 g_R^2 \, \chi_R^2 \, \left(\log \frac{\chi_R^2}{\chi_0^2} - 3 \right)
\end{equation}
Taking $\partial V_{GE} / \partial \chi_R = 0$, we see that there is a local maximum at $\chi_R = 0$ and two global minima at
\begin{equation}
\label{finalGap1}
	\frac{\chi_R}{e \chi_0} = \pm \exp \left[ - \frac{2\pi}{(2N - 1) g_R^2} \right]
\end{equation}
It is obvious that the dependence on the renormalized coupling $g_R$ is of a form that is not accessible via perturbation theory.  Since $\chi_R = 0$ is a local maximum of the renormalized Gaussian effective potential, it does not correspond to a stable vacuum.  On the other hand, the two extrema in Eqn. (\ref{finalGap1}) correspond to global minima, and hence vacua of the theory.  Since these global minima occur at non-zero values of $\chi$, and thus non-zero values of $m$, we have dynamical mass generation in the Gross-Neveu model.   As remarked before, dynamical mass generation in the Gross-Neveu model is well known, and our result establishes it within the Gaussian variational approximation.  The renormalized Gaussian effective potential as a function of the renormalized field $\chi_R$ is shown in the following figure.

\begin{figure}[hp]
	\begin{centering}
	\scalebox{1.0}{\includegraphics[width=1.0\textwidth,center]{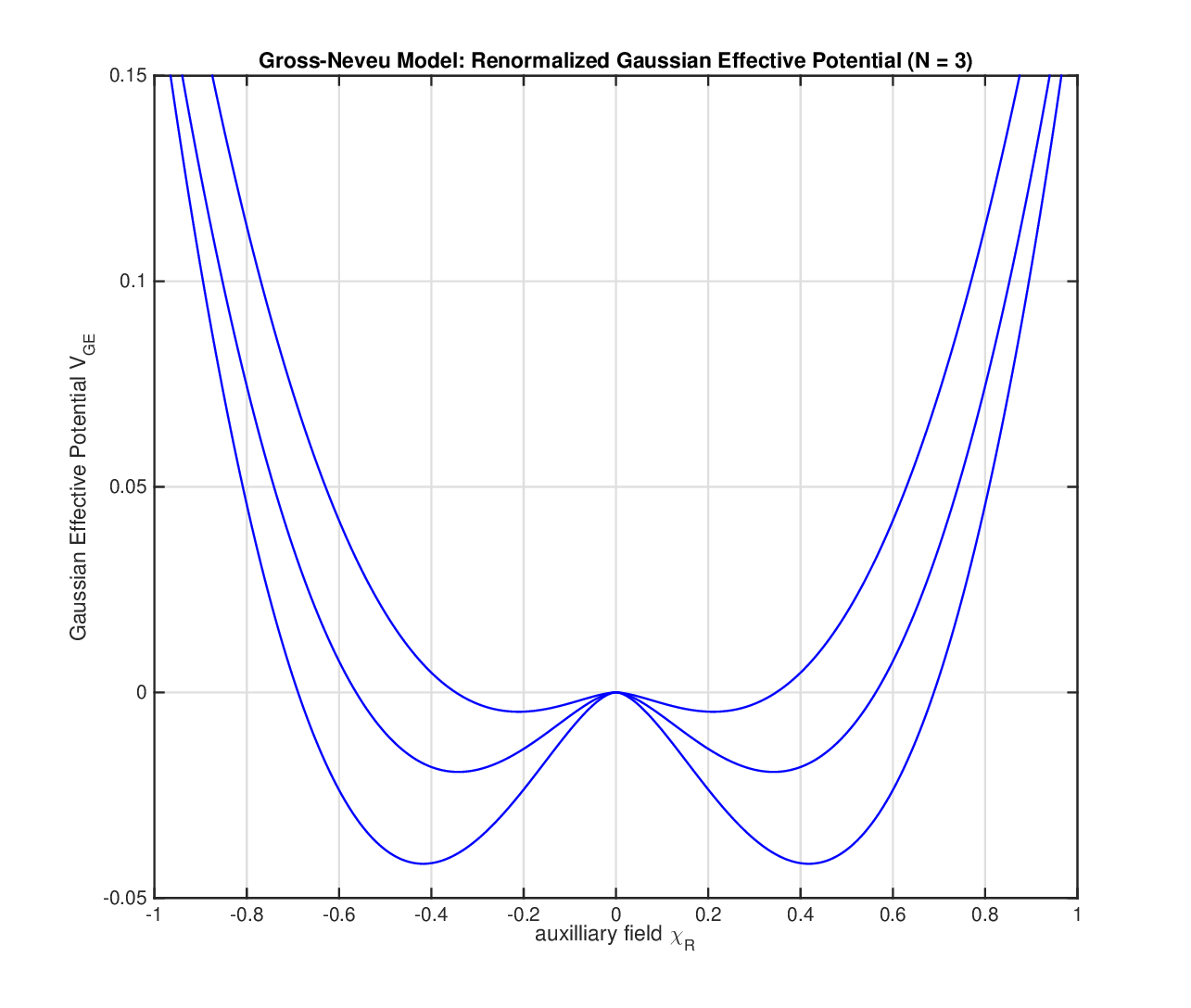}}
	\end{centering}
	\caption{Renormalized Gaussian effective potential for different values of the renormalized coupling, at a fixed value of $N = 3$.  Curves from upper to lower have renormalized couplings $g_R = 0.8$, $g_R = 1.0$, and $g_R = 1.2$, respectively.  }
	\label{fig:gep_gn}
\end{figure}

\section{Entanglement Entropy Calculations for the Gross-Neveu Model}
\label{sec:vcee0}

In this section, we will compute the variational approximation to the entanglement entropy of the Gross-Neveu model on the half-line.  Recall that the  entanglement entropy for the ground state of $N$ free fermions field of mass $m$ in $1+1$ dimensions on the half-line is \cite{Solo1}
\begin{equation}
\label{EE1}
	S = - \frac{N}{12} \left(\log \varepsilon^2 + \log m^2 \right) + \mathcal{O}(m^2 \varepsilon^2)
\end{equation} 
where $\varepsilon$ is our short-distance cutoff.  We will ignore the $\mathcal{O}(m^2 \varepsilon^2)$ corrections which go to zero as $\varepsilon \to 0$.  The free fermion entanglement entropy can be derived in a wide variety of ways, including using the heat kernel method.

Our Gaussian variational ansatz is identical in form to the ground state of a free fermionic field, but with an $m$ given by the variational mass.  First, we will examine the entanglement entropy for Gaussian states with fixed $\chi_R$, and then we will consider the entanglement entropy in the ground state of Gross-Neveu model.

\subsection{Entanglement Entropy for fixed $\chi_R$}
\label{sec:vcee1}

In the variational approximation, we find that for fixed $\chi_R$ the Gross-Neveu entanglement entropy on the half-line is simply
\begin{equation}
\label{EE1}
	S = - \frac{N}{12} \left[\log(\varepsilon^2) + \log \left(\left(\frac{2N-1}{2N} \right)^2\, g_R^2 \chi_R^2 \right) \right] 
	+ \mathcal{O}(m^2 \varepsilon^2)
\end{equation} 
where we have used the renormalized gap equation (\ref{eq:renormalized_gap_equation}).  It follows immediately that $S$ is monotonically decreasing and convex as a function of the coupling $g_R$ for $g_R > 0$.  In particular, for $g_R > 0$ we have
\begin{align}
	\frac{\partial S}{\partial g_R} &= - \frac{N}{6 g_R}< 0 \\ \nonumber \\
	\frac{\partial^2 S}{\partial g_R^2} &=  \frac{N}{6 g_R^2} > 0
\end{align}
which are both cutoff independent.  The monotonicity and convexity of $S$ with respect to $g_R$ on the half-line exhibit the same behavior as the variational approximation to the entanglement entropy of scalar $\phi^4$ theory on the half-line, which is likewise monotonically decreasing and convex with respect to its coupling \cite{CotlerMueller1}.  In both cases, the behavior is intuitive: increasing the coupling localizes correlations, which decreases entanglement.  Since the von Neumann entropy is bounded from below, for a fixed cutoff the von Neumann entropy can only decrease by a finite amount as we arbitrarily increase the coupling.  Thus, we expect the von Neumann entropy to be convex with respect to the coupling.

We can also examine the behavior of $S$ with respect to $N$.  Indeed, $S$ is monotonically increasing and concave with respect to $N$ for $N \geq 1$, since
\begin{align}
	\label{N1}
	\frac{\partial S}{\partial N} &= \frac{S}{N} - \frac{1}{6(2N - 1)} > 0 \\ \nonumber \\
	\label{N2}
	\frac{\partial^2 S}{\partial N^2} &= - \frac{S}{N^2} + \frac{1}{6N(2N-1)^2} < 0\,.
\end{align}
Monotonicity is dominated by the $- \log(\varepsilon)$ divergence, which is positive infinity.  Similarly, concavity is dominated by $\log(\varepsilon)$, which is negative infinity.

One might be hesitant to trust the variational approximation to provide us information about the $N$-dependence of $S$.  We expect the class of states under consideration to look approximately like the ground state of a free fermionic field for small coupling $g_R$.  However, we do not a priori expect the variational approximation to be accurate for arbitrary $N$ and fixed, small $g_R$.  In particular, the variational approximation appears sensible when $g_R^2 N$ is small.  So while the $N$-dependence of the variational approximation to the entanglement entropy is interesting, we should be cautious about its interpretation.

In the next section, we examine the entanglement entropy of the Gross-Neveu model in its ground state.

\subsection{Entanglement Entropy in the Ground State}
\label{sec:vcee2}

There are two vacuum states in the Gaussian approximation of the Gross-Neveu model located at
\begin{equation}
\label{vacua1}
	\frac{\chi_R}{e \chi_0} = \pm \exp \left[ - \frac{2\pi}{(2N - 1) g_R^2} \right]
\end{equation}
Since the theory lives in 1+1 dimensions, tunneling between the vacua can occur.  Without getting into details here, we refer the reader to the original paper of Gross and Neveu \cite{GrossNeveu1} for a discussion of the meaning of spontaneous symmetry breaking in the model, and the implications for dynamical mass generation.

Let us consider the entanglement entropy of each vacuum state separately.  We note that the entanglement entropy in Eqn. (\ref{EE1}) is a function of $\chi_R^2$, and so Eqn. (\ref{vacua1}) tells us that each of the vacua has the same entanglement entropy:
\begin{equation}
	S_{\text{gs}} = - \frac{N}{12} \left[2 \log \varepsilon^2 + \log\left(\left(\frac{2N-1}{N}\right)^2 \, g_R^2 \chi_0^2 \right) 
	+ \left(2 - \frac{4 \pi}{(2N-1) g_R^2} \right) \right]
\end{equation}
The $1/g_{R}^2$ term in the above formula is a hallmark of non-perturbative behavior. We have that $S_{\text{gs}}$ is monotonically decreasing and convex as a function of the coupling $g_R$ for $g_R > 0$.  For $g_R > 0$, we have
\begin{align}
	\frac{\partial S_{\text{gs}}}{\partial g_R} &= - \frac{N}{6 g_R} - \frac{2\pi N}{3 (2N-1) g_R^3} < 0 \\ \nonumber \\
	\frac{\partial^2 S_{\text{gs}}}{\partial g_R^2} &=  \frac{N}{6 g_R^2} + \frac{2\pi N}{(2N-1) g_R^4} > 0
\end{align}
which are cutoff independent quantities.

We will now analyze the behavior of $S_{\text{gs}}$ with respect to the number of colors $N$.  As before, $S_{\text{gs}}$ is monotonically increasing and concave with respect to $N$ for $N \geq 1$, since
\begin{align}
	\label{N1}
	\frac{\partial S_{\text{gs}}}{\partial N} &= \frac{S_{\text{gs}}}{N} - \frac{1}{6(2N-1)} - \frac{2 \pi N}{3 (2N-1)^2 g_R^2}> 0 \\ \nonumber \\
	\label{N2}
	\frac{\partial^2 S_{\text{gs}}}{\partial N^2} &= - \frac{S_{\text{gs}}}{N^2} + \frac{1}{6N(2N-1)^2}  + \frac{4 \pi}{3 (2N - 1)^3 g_R^2}< 0\,.
\end{align}
Here we have the usual caveat about the accuracy of the variational approximation as a function of $N$, as well as the divergence of $\pm \log(\varepsilon)$.

\section{Renormalization Group and Entanglement Entropy}
\label{sec:rgee}

From Eqn. (\ref{EE1}), it follows that the renormalization group equation for the entanglement entropy is
\begin{equation}
	\label{EEbeta1}
	\mu \,\frac{dS}{d\mu} = - \frac{N}{6 m} \frac{dm}{dg_R} \,\beta(g_R)
\end{equation}
where $\beta(g_R)$ is the beta function for the renormalized coupling.  We analyze the renormalization group equation for the entanglement entropy of the Gross-Neveu model for Gaussian states with fixed $\chi_R$, and in the variational approximation of the ground state vacua.

\subsection{Renormalization Group for Entanglement Entropy at fixed $\chi_R$}
\label{ssec:rggn1}

For the Gross-Neveu model, we have
\begin{equation}
	\label{EEBeta2}
	\mu \,\frac{dS}{d\mu} = - \frac{N}{6 m} \frac{d m}{dg_R} \,\beta(g_R)
\end{equation}
and the renormalized gap equation (\ref{eq:renormalized_gap_equation}) for fixed $\chi_R$ gives us
\begin{equation}
	\frac{dm}{dg_R} =  \frac{m}{g_R}
\end{equation}
To obtain the Gross-Neveu beta function, we differentiate the coupling renormalization equation \eqref{eq:coupling_renormalization} with respect to the renormalization scale $\mu$ at fixed bare coupling and fixed UV cutoff (in $I_0(\mu)$) and find 
\begin{equation}
	\beta(g_R) = \mu \, \frac{d g_R}{d \mu} = - \frac{2N - 1}{4\pi} \, g_R^3
\end{equation}
We see that the theory is asymptotically free.  Furthermore, Eqn. (\ref{EEBeta2}) becomes
\begin{equation}
\label{EEBeta3}
\mu \,\frac{dS}{d\mu} = \frac{N(2N - 1)}{24\pi} \, g_R^2
\end{equation}
which is not itself ``asymptotically free."  That is, unlike the renormalized coupling $g_R$ which is \emph{decreasing} at higher renormalization mass scales $\mu$, we find that the entanglement entropy $S$ is \emph{increasing}, albeit less and less rapidly as $\mu$ increases because of the asymptotically free coupling.

\subsection{Renormalization Group for Entanglement Entropy in the Ground State}
\label{ssec:rggn2}

Recall that there are two vacua for the Gross-Neveu model given by Eqn. (\ref{vacua1}).  In each of these vacua, we use Eqn.'s (\ref{eq:renormalized_gap_equation}) and (\ref{vacua1}) to find that
\begin{equation}
	\frac{d m}{d g_R} = \left(\frac{1}{g_R} + \frac{4 \pi}{(2N-1) g_R^3} \right) \, m
\end{equation}
and so
\begin{equation}
	\label{EEBeta4}
	\mu \,\frac{dS_{\text{gs}}}{d\mu} = \frac{N(2N-1)}{24 \pi} \, g_R^2 + \frac{N}{6}
\end{equation}
which is similarly not asymptotically free.  So, as was the case in the previous subsection where we considered a state with arbitrary $\chi_R$, we find that the entanglement entropy for either vacuum state is also \emph{increasing} as the renormalization mass scale $\mu$ gets larger.  Interestingly, taking $g_R \to 0$ we see that the right hand side of Eqn. (\ref{EEBeta4}) becomes the constant $N/6$.

\section{Conclusions}
\label{sec:conclusions}

We have computed variational approximations to the entanglement entropies of vacua and more general coherent states with arbitrary values of $\chi_R$ in the Gross-Neveu model.  The entanglement entropies are monotonically decreasing and convex as a function of the coupling, which is the same as the behavior of the entanglement entropy in $\phi^4$ theory in 1+1 dimensions \cite{CotlerMueller1}.  This agreement provides evidence for the conjecture that entanglement entropies for relativistic field theories in 1+1 dimensions are generically monotonically decreasing and convex in their couplings \cite{CotlerMueller1}.  Additionally, we have shown that within the variational approximation, the entanglement entropies are monotonically increasing and concave in the number of colors $N$.

Although the Gross-Neveu model is asymptotically free, its entanglement entropy is not asymptotically free for the states we studied.  Interestingly, the beta function of the entanglement entropy of the vacua contains a constant term which indicates that even for vanishing renormalized coupling, there is singular behavior when the renormalization group flow is integrated.  This behavior would be interesting to explore further.

The only other example of an asymptotically free theory for which we can variationally compute the entanglement entropy is precarious $\phi^4$ theory in 3+1 dimensions \cite{CotlerMueller1}.  In this theory, the beta function of the entanglement entropy is asymptotically free, in contrast with our results for the Gross-Neveu model.  One might expect that the entanglement entropies of asymptotically free theories should have similar behavior of their renormalization group flows, but apparently this is not the case.

However, we note that the entanglement entropy in precarious $\phi^4$ theory is monotonically \textit{increasing} in the coupling strength -- that is, in the absolute value of the renormalized coupling \cite{CotlerMueller1}.  Perhaps this distinguishing feature is responsible for the different behavior of the renormalization group flow of precarious $\phi^4$ theory vis-\`{a}-vis the Gross-Neveu model.  This hypothesis seems ripe for future exploration.

The variational approach presented in this paper provides useful tools for analyzing non-perturbative features of fermionic field theories, including entanglement entropies.  These tools can be effectively applied to fermionic theories beyond the Gross-Neveu model, as well as fermionic theories coupled to bosonic theories via techniques presented in \cite{CotlerMueller1}.  Further applications promise to broaden our understanding of entanglement entropy in a wide variety of coupled field theories for which standard techniques are inapplicable.

\newpage
\section{Appendix: Clifford Algebra and Dirac Spinors}

\paragraph{Dirac spinor fields $\psi_{\alpha}^a(\mathbf{x})$.}  In both 1+1 and 2+1 dimensions the spinor indices $\{\alpha, \beta, \ldots\}$ take the values $\{1, 2\}$ and the color indices $\{a, b, \ldots\}$ take the values $\{1, \ldots, N\}$.  The hermitian conjugate spinor is denoted $\psi_{\alpha}^{a \dagger}$ and the Dirac conjugate spinor is $\bar{\psi}_{\beta}^a = \psi_{\alpha}^{a \dagger} \beta_{\alpha\beta}$.

\paragraph{Clifford algebra in $1+1$ and $2+1$ dimensions.}  Spacetime indices will be denoted $\mu, \nu, \ldots$, and spatial indices $m, n, \ldots$, ranging over the obvious values in each dimension.
\begin{equation}
	\gamma^{\mu} \gamma^{\nu} + \gamma^{\nu} \gamma^{\mu} = - 2 \eta^{\mu\nu}
\end{equation}
where we use a metric with $\eta^{00} = -1$ and $\eta^{mm} = 1$.  In $1+1$ dimensions we can define $\gamma^2 = -i \gamma^0 \gamma^1$ which satisfies $(\gamma^2)^2 = -1$, and in $2+1$ dimensions $\gamma^2$ is considered to be part of the basis for the Clifford algebra.  In the Hamiltonian formulation, we need the matrices $\alpha^{\mu}$ and $\beta$ defined by
\begin{align}
	\alpha^{\mu} &= \gamma^0 \gamma^{\mu}  \\
	\beta &= \gamma^0
\end{align}
so $\alpha^0 = (\gamma^0)^2 = 1$ is the identity.  The $\alpha^m (= \alpha_m)$ and $\beta$ matrices satisfy
\begin{align}
	\alpha^m \alpha^n &+ \alpha^n \alpha^m = 2 \delta^{mn}  \\
	\alpha^m \beta &+ \beta \alpha^m = 0  \\
	\beta^2 &= 1
\end{align}
Whether we are in spacetime dimension 1+1 or 2+1, the representations of the Clifford algebra are two-dimensional.  For detailed calculations, we use a Dirac representation of the Clifford algebra (where $\beta = \gamma^0$ is diagonal).  This two-dimensional representation is given in terms of the standard Pauli matrices $\sigma_i$ with indices $\{i, j, \ldots\}$ taking the values $\{1, 2, 3 \}$, by
\begin{equation}
	\alpha^1 = \sigma_1  \quad\quad\quad
	\alpha^2 = \sigma_2  \quad\quad\quad
	\beta = \sigma_3
\end{equation}
or in terms of the gamma-matrices
\begin{equation}
	\gamma^0 = \sigma_3  \quad\quad\quad
	\gamma^1 = i \sigma_2 \quad\quad\quad
	\gamma^2 = -i \sigma_1
\end{equation}

\newpage
\section*{Acknowledgments}
\label{sec:ack}
The authors would like to thank Roman Jackiw, Paul Stevenson, and Frank Wilczek for valuable discussions, and Jae-Hyung Yee for helpful correspondence and providing a copy of his review article.  Jordan Cotler is supported by the Fannie and John Hertz Foundation and the Stanford Graduate Fellowship program.  Mark Mueller is supported by the U.S. Department of Energy under grant Contract Number DE-SC00012567.

\end{document}